\documentclass[journal=jacsat,manuscript=article]{achemso}
\setkeys{acs}{doi = true}

\usepackage[version=3]{mhchem}
\usepackage{siunitx}
\usepackage{amssymb}
\usepackage{bm}
\usepackage{placeins}
\usepackage{float}
\usepackage{lineno}
\usepackage{glossaries}
\usepackage{booktabs}
\usepackage{graphicx}
\usepackage{etoolbox}
\usepackage{multirow}
\robustify\itshape
\usepackage{nth}

\DeclareSIUnit{\mAh}{mAh}
\DeclareSIUnit{\watthour}{Wh}
\DeclareSIUnit{\kWh}{kWh}
\DeclareSIUnit{\million}{M}
\DeclareSIUnit{\COtwoeq}{\ensuremath{\mathrm{CO_2eq}}}
\newcommand{\numinterval}[2]{[\num{#1}, \num{#2}]}
\usepackage{color,soul}
\usepackage{xcolor}
\usepackage[colorlinks = true,
            linkcolor = blue,
            urlcolor  = blue,
            citecolor = blue,
            anchorcolor = blue]{hyperref}
\usepackage{makecell}

\newcommand{\doi}[1]{\href{http://dx.doi.org/#1}{\nolinkurl{#1}}}

\usepackage{etoolbox}

\makeatletter
\renewcommand*{\acs@author@fnsymbol@symbol}[1]{
    \ifcase #1 *\or
    1\or
    2\or
    3\or
    4\or
    5\or
    6\or
    7\or
    8\or
    9\or
    10
    \fi
}
\renewcommand*\acs@contact@details{%
    {\sffamily *\,E-mail: \acs@email@list }%
    \acs@number@list
}           
\patchcmd{\acs@address@list@auxii}%
{\acs@author@fnsymbol{\acs@affil@marker@cnt}}
{\textsuperscript{\acs@author@fnsymbol{\acs@affil@marker@cnt}}}
{}{}
\patchcmd{\acs@address@list@auxii}%
{{\acs@author@fnsymbol{\acs@affil@marker@cnt}\@nameuse{@altaffil@\@roman\@tempcnta}\par}}
{{\textsuperscript{\acs@author@fnsymbol{\acs@affil@marker@cnt}}\@nameuse{@altaffil@\@roman\@tempcnta}\par}}
{}{}        
\makeatother    

\author{Subhash V.S. Ganti}
\affiliation[1]{University of Bayreuth, Faculty of Engineering Science, Universitätsstraße 30, 95447 Bayreuth, Germany}
\alsoaffiliation[2]{University of Bayreuth, Bavarian Center for Battery Technology (BayBatt), Universitätsstraße 30, 95447 Bayreuth, Germany}
\author{Lukas Wölfel}
\affiliation[1]{University of Bayreuth, Faculty of Engineering Science, Universitätsstraße 30, 95447 Bayreuth, Germany}
\author{Arla Diengdoh}
\affiliation[1]{University of Bayreuth, Faculty of Engineering Science, Universitätsstraße 30, 95447 Bayreuth, Germany}
\author{Tzvetan Mihaylov}
\affiliation[3]{Digital R\&D, Covestro Deutschland AG, Kaiser-Wilhelm-Allee 60, 51373 Leverkusen, Germany}
\author{Thomas Nevolianis}
\affiliation[3]{Digital R\&D, Covestro Deutschland AG, Kaiser-Wilhelm-Allee 60, 51373 Leverkusen, Germany}
\author{Jan Meyer}
\affiliation[3]{Digital R\&D, Covestro Deutschland AG, Kaiser-Wilhelm-Allee 60, 51373 Leverkusen, Germany}
\author{Christopher Kuenneth}
\affiliation[1]{University of Bayreuth, Faculty of Engineering Science, Universitätsstraße 30, 95447 Bayreuth, Germany}
\alsoaffiliation[2]{University of Bayreuth, Bavarian Center for Battery Technology (BayBatt), Universitätsstraße 30, 95447 Bayreuth, Germany}

\email{christopher.kuenneth@uni-bayreuth.de}

\title[]{HorizON: A Multi-Horizon Deep Operator Network for Property Prediction of Organic Battery Stacks}

\begin{document}

\begin{abstract}
Lithium-ion batteries based on transition-metal oxide cathodes depend on critical elements such as cobalt and nickel, whose extraction raises environmental and supply-chain concerns. Organic electrode materials (OEMs) offer a promising alternative because their versatile structures enable systematic tuning of electrochemical properties. However, OEM based batteries still suffer from limited cycling stability due to challenges such as active material dissolution into the electrolyte, side reactions, and limited electronic conductivity. Addressing these challenges requires moving beyond the active material alone and considering the complete battery stack and the coupled interactions among them. This complexity requires the joint optimization of multiple properties, including voltage, gravimetric capacity, volumetric capacity, energy density, and capacity retention. Machine learning accelerates this optimization by learning structure-property relationships across battery components and enables battery material discovery. However, most existing models predict instantaneous properties and therefore provide limited insight into how battery properties evolve over cycles. A predictive framework capable of learning across multiple properties, data fidelities, and prediction horizons is thus needed. Here, we introduce HorizON, a multi-fidelity, multi-horizon deep operator network that predicts battery properties such as specific capacity versus cycle number and average voltage for organic battery stacks. HorizON takes fingerprints of the complete materials stack as inputs, along with usage conditions such as C-rate, mass loading, and voltage window. The model is trained jointly on DFT-derived and experimental data, exploiting correlations across properties and fidelity levels. On the test dataset, HorizON achieves an R2 (RMSE) of 0.96 (\SI{24.6}{\milli\ampere\hour\per\gram}) for capacity and 0.97 (\SI{0.22}{\volt}) for voltage over cycles. We then deploy HorizON for large scale discovery and optimization of polymer cathodes. Here, HorizON is utilized to predict voltage and specific capacity for libraries exceeding 100 million chemically valid polymers spanning multiple polymer classes, with inference time below \SI{1}{\milli\second} per candidate. This capability enables exploration of chemical spaces that are difficult to investigate using conventional simulations or experiments. By identifying promising materials and suitable usage conditions, HorizON can help experimentalists move from trial-and-error synthesis and cell testing toward targeted, hypothesis driven experiments. We further apply inverse design to identify polymers with targeted battery properties and process optimization to identify suitable battery components and usage conditions. We also apply explainable artificial intelligence to identify chemical fragments associated with improved battery performance. These relationships provide a molecular interpretation of the learned structure-property relationships and connect predicted battery behavior to chemically meaningful polymer features. Beyond organic batteries, the framework is transferable to inorganic, redox-flow, and other battery chemistries, offering a general route for data-driven discovery in complex materials systems with multiple interacting components.
\end{abstract}

\begin{figure}[!htbp]
\centering
\includegraphics{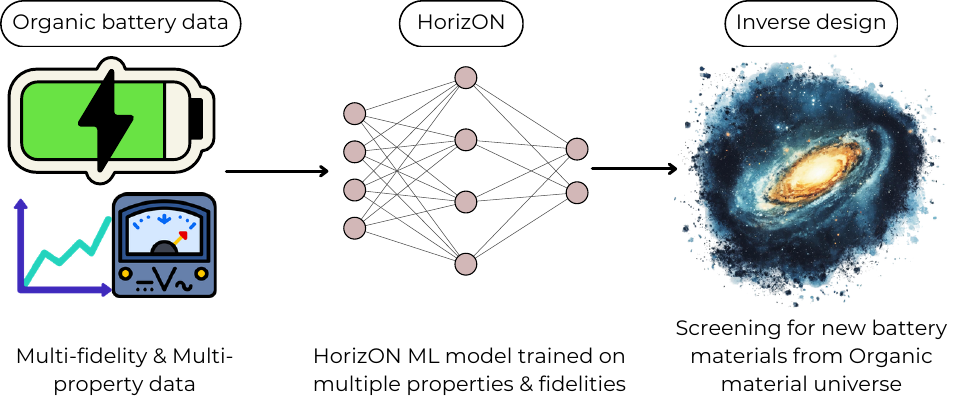}
\end{figure}
\FloatBarrier

\clearpage
\section{Introduction}
Organic electrode materials (OEMs) are promising alternatives to traditional inorganic electrodes such as lithium, cobalt and nickel. \cite{lu_prospects_2020,kim_organic_2022,larcher_towards_2015} OEMs are sustainable, highly abundant and exhibit high structural flexibility.\cite{shi_challenges_2022} Moreover, they are not prone to supply-chain risks and pose little negative environmental impact.\cite{lu_prospects_2020} Owing to their tunable redox-active chemistries, they have shown electrochemical viability in lithium-ion cathodes (\qtyrange{3.0}{4.5}{\volt}), sodium-ion cells (\qtyrange{2.5}{3.5}{\volt}), multivalent Mg$^{2+}$/Zn$^{2+}$/Al$^{3+}$ chemistries, aqueous systems, and solid-state configurations. \cite{wang_conjugated_2021,judez_energy_2019,bertaglia_eco-friendly_2024,lee_recent_2018,esser_perspective_2021} To date, roughly \num{200} organic materials, including carbonyls, phenolics, viologens, and conjugated polymers, have been reported as electroactive battery materials.\cite{schon_rise_2016,liang_organic_2012,bhosale_organic_2018}. However, many existing OEMs exhibit capacity loss after a few hundreds of cycles.\cite{shi_challenges_2022} The capacity loss is primarily due to the dissolution of active species into the electrolyte, side reactions, and low intrinsic electronic conductivity.\cite{shi_challenges_2022,lee_recent_2018,yu_review_2024} Overcoming these limitations require the design of new OEMs across a large possible chemical spaces.\cite{reymond_chemical_2015,lu_perspectives_2023} 

Machine learning (ML) provides an efficient approach for navigating this large chemical space and predicting battery properties.\cite{butler_machine_2018,shen_machine_2022,chen_machine_2020} ML models learn structure--property relationships from available battery data.\cite{wu_moleculenet_2018} For OEMs, Simplified Molecular Input Line Entry System (SMILES) strings are commonly used to represent their chemical structures.\cite{weininger_smiles_1988} These strings are converted into machine-readable chemical fingerprints using methods such as polyBERT\cite{kuenneth_polybert_2023}, Mordred\cite{moriwaki_mordred_2018}, or circular fingerprints\cite{rogers_extended-connectivity_2010}. The resulting fingerprints serve as model inputs for predicting battery properties such as specific capacity and voltage.\cite{Ganti2025} Moreover, data-fusion and multi-task ML based methods enable joint learning across multiple battery properties and data sources.\cite{hutchinson_overcoming_2017,caruana_multitask_1997,Nevolianis2025b} In our previous work, we used a multi-task ML model to predict first-cycle specific capacity and voltage for different cathode--anode pairs across several organic material classes.\cite{Ganti2025}

However, practical battery applications require predictions of cycling stability beyond first-cycle performance.\cite{lu_prospects_2020,lee_recent_2018,schon_rise_2016} The cycling stability of OEMs is particularly limited by the active-material dissolution and side reactions.\cite{lu_prospects_2020,lee_recent_2018,shi_challenges_2022} Several strategies can mitigate these issues: (i) crosslinking to suppress dissolution; (ii) polar or ionic substituents or salt formation to reduce solubility; and (iii) electron-withdrawing or electron-donating substituents to tune the redox potential and operating voltage.\cite{shi_challenges_2022,lee_recent_2018,schon_rise_2016} However, each of these strategies demand extensive optimization of the redox-active substituent, its position, and the battery materials stack design, which makes experimental screening slow.\cite{schon_rise_2016,lee_recent_2018} Recent studies have applied sequence models, including RNNs, LSTMs, and attention-based temporal fusion transformers, to predict long-term capacity fade, voltage evolution, and remaining useful life.\cite{Zhang2018,lim_tft_2021,thakkar_advances_2024} However, these recent studies only focused on a single chemistry or cell design, limiting their transferability across OEMs.\cite{severson_data-driven_2019,thakkar_advances_2024} Achieving such transferability requires a model that captures both the atomistic chemistry of OEMs and information from the full battery materials stack. We therefore develop a multi-horizon, multi-fidelity, and multi-property ML model that integrates these inputs to predict long-term performance across diverse OEM chemistries.

\begin{figure}[p]
\centering
\includegraphics{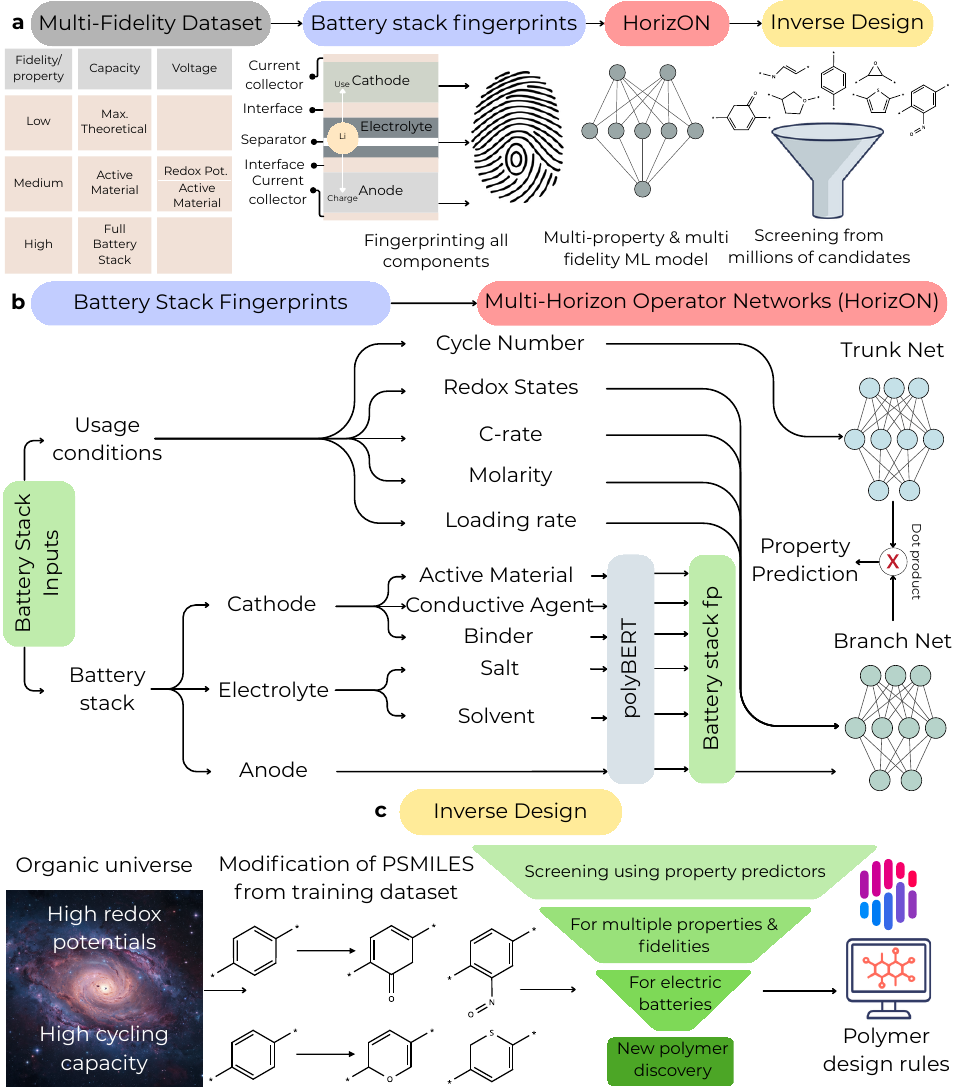}
\caption{\textbf{Multi-Horizon Deep Operator Network workflow.} \textbf{a} The four stages are multi-fidelity dataset construction, battery-stack fingerprinting, Multi-Horizon Deep Operator Network modeling, and inverse design. We encode the cell components as polyBERT fingerprints and combine them with fidelity and usage condition descriptors for multi-property prediction. \textbf{b} The Multi-Horizon Deep Operator Network architecture provides the cycle number to the trunk network and the usage conditions and battery-stack fingerprints to the branch network. A dot product combines the two latent representations to predict the target property. \textbf{c} The inverse-design workflow iteratively modifies the chemical structure of training polymers, screens the generated candidates with the trained model, and uses SHapley Additive exPlanations (SHAP) to derive interpretable polymer design rules.}
\label{fig:workflow}
\end{figure}

Figure~\ref{fig:workflow}\textbf{a} summarizes the HorizON pipeline for heterogeneous electrochemical data. Our previous multi-task model predicted first-cycle voltage and specific capacity across diverse organic battery chemistries~\cite{Ganti2025}. We integrate the corresponding voltage and capacity data as the medium-fidelity subset because it excludes electrolyte and usage condition descriptors. HorizON combines these data with high-fidelity experimental capacity-versus-cycle trajectories and medium-fidelity adiabatic DFT redox potentials versus the standard hydrogen electrode (SHE). The high-fidelity trajectories include the complete organic battery-stack composition and usage conditions. Our HorizON approach comprises of five models trained cross-validated subset and a meta-learner ensemble that combines their predictions. Using the homopolymer PSMILES (Polymer Simplified Molecular Input Line Entry System) strings in the training dataset as reference structures, we generate a library of four million polymers through iterative substitutions and additions of redox-active fragments, then screen the candidates for improved cycling stability and redox potential. Finally, SHapley Additive exPlanations (SHAP) identify preferred redox-active moieties and provide interpretable design rules.\cite{Lundberg2017}

\section{Results and Discussion}

\begin{table}[t]
\centering
\footnotesize
\setlength{\tabcolsep}{6pt}
\renewcommand{\arraystretch}{1.2}
\caption{Overview of the multi-fidelity dataset used to train the model. We combine high-fidelity cycling data with medium-fidelity active-material experimental measurements, medium-fidelity DFT redox potentials, and low-fidelity theoretical capacities. Capacities are reported in \si{\mAh\per\gram} and voltages/potentials in \si{\volt}. An em-dash (---) marks quantities not collected for that subset.}
\label{tab:dataset}
\begin{tabular}{@{}llclrrrcc@{}}
\toprule
Property & Subset & Cycle & Method & Data pts. & \# Cathodes & \# Electrolytes & Range & Fidelity \\
\midrule
\multirow{3}{*}{Capacity}
 & Full-stack\textsuperscript{a}      & many & Exp.\           & 457\textsuperscript{a} & 18  & 6   & \numinterval{11.0}{223.0}  & High \\
 & Active material\textsuperscript{b} & 1st   & Semi-Exp.\textsuperscript{c} & 217 & 217 & --- & \numinterval{8.0}{691.0}   & Medium \\
 & Theoretical\textsuperscript{b}     & ---   & Faraday         & 130 & 130 & --- & \numinterval{93.0}{570.0}  & Low \\
\midrule
\multirow{2}{*}{Voltage}
 & Active material\textsuperscript{b,d} & 1st   & Semi-Exp.\textsuperscript{c} & 197 & 197 & --- & \numinterval{0.6}{4.07}    & Medium \\
 & Redox potential\textsuperscript{d}                    & ---   & DFT             & 67\textsuperscript{e}  & 45\textsuperscript{e}  & --- & \numinterval{-3.2}{5.45}   & Medium \\
\midrule
\multicolumn{4}{@{}r}{Total} & 1068 & 607 & 6 & & \\
\bottomrule
\end{tabular}

\vspace{2pt}
\begin{flushleft}\footnotesize
\textsuperscript{a}\,Each battery stack comprises a cathode, lithium-metal anode, and electrolyte and contributes a distinct number of measured cycles. Together, these cycle-resolved measurements yield \num{457} data points.\par
\textsuperscript{b}\,Organic cathode data reused from our previous publication \cite{Ganti2025}.\par
\textsuperscript{c}\,Semi-Exp.: experimental values using only the active material in the cathode and the anode (lithium). Usage conditions and electrolyte information are not available.\par
\textsuperscript{d}\,Active-material voltage is referenced to \ce{Li/Li+}; redox potential to SHE.\par
\textsuperscript{e}\,The \num{67} redox-potential data points correspond to \num{45} unique cathode structures because some organic materials contribute multiple oxidation- or reduction-state transitions. The associated oxidation and reduction states are encoded as usage conditions in selector vectors.
\end{flushleft}
\end{table}
We collect and curate a dataset of \num{1068} data points for organic cathode materials across three fidelity levels. A synopsis of this dataset can be found in Table~\ref{tab:dataset}. Experimental measurements are collected from peer-reviewed literature\cite{Otteny2018,Bhosale2024,Obrezkov2021,Dai2019,Xu2021,Liu2021,Peterson2018,Iordache2016}, while simulated redox potentials are generated using DFT calculations (see the Methods section for details). The high-fidelity subset covers organic battery stacks containing \num{18} unique polymer cathodes, lithium-metal anodes, and six different electrolytes. These stacks contribute \num{457} experimental capacity-versus-cycle measurements, corresponding to \qty{43}{\percent} of the dataset, each including the complete battery-stack composition and usage conditions. For each measurement, we document the active-material loading (\qtyrange{0.45}{2.5}{\milli\gram\per\centi\meter\squared}), the voltage window (\qtyrange{1.5}{4.5}{\volt}), and the C-rate (\numrange{10}{1000}), followed by the experimentally measured specific capacities (\qtyrange{11}{223}{\mAh\per\gram}).

The medium-fidelity subset contains both experimental single-cycle data, reused from our previous publication \cite{Ganti2025}, and simulated DFT redox potentials. It contains \num{414} points, corresponding to \qty{39}{\percent} of the dataset, and covers \num{259} homopolymer cathodes measured against a lithium anode. This subset includes \num{217} first-cycle discharge capacities (\qtyrange{8.0}{691}{\mAh\per\gram}) and \num{197} average voltages (\qtyrange{0.6}{4.07}{\volt}), both measured versus lithium. The DFT dataset contains \num{67} simulated redox potentials computed for \num{45} homopolymer structures at different oxidation and reduction states. The potentials are referenced to the standard hydrogen electrode (SHE) and span \qtyrange{-3.2}{5.45}{\volt}. The number of potentials exceeds the number of structures because some structures exhibit multiple redox-state transitions. The low-fidelity subset reuses theoretical-capacity data from our previous publication \cite{Ganti2025}. This subset contains \num{130} computed theoretical capacities for the same homopolymer cathodes and spans \qtyrange{93.0}{570.0}{\mAh\per\gram}.

\paragraph{Battery materials stack fingerprinting}
As illustrated in Figure~\ref{fig:workflow}, HorizON receives fingerprints of the complete battery materials stack as inputs. We first generate a \num{600}-dimensional polyBERT fingerprint for each material as indicated in Figure~\ref{fig:workflow} .\cite{kuenneth_polybert_2023} For each battery materials stack, we calculate a composition-weighted fingerprint by summing the constituent fingerprints weighted by their respective fractions\cite{kuenneth_copolymer_2021}. The cathode fingerprint is built from the OEM active material, conductive agent, and binder, whereas the electrolyte fingerprint is created using salt and solvent. The lithium-metal anode is represented by a separate \num{600}-dimensional fingerprint. We concatenate the cathode, electrolyte, and anode fingerprints with seven selector variables describing the usage conditions and redox states. The resulting \num{1807}-dimensional input enables HorizON to learn interactions across the complete battery materials stack.

\paragraph{Multi-horizon deep operator network (HorizON)}
Building on the complete battery-stack representation, HorizON uses a deep operator network to predict both cycle-dependent and scalar battery properties. The branch network receives the \num{1807}-dimensional input containing the battery-stack fingerprints, usage conditions, and redox-state descriptors as described in the previous section. The trunk network receives the cycle number. Each network maps its inputs to a latent space representation, and their dot product is passed through fully connected layers to predict the target voltage and specific capacity. Evaluating the trunk network at multiple cycle numbers produces a complete specific-capacity trajectory for a fixed battery stack and set of usage conditions. For scalar targets without cycle dependence, such as DFT redox potentials, we set the trunk input to cycle \num{1}. The shared architecture is trained jointly across target properties and fidelity levels, enabling multi-property and multi-fidelity learning over cycles within a single model.

\begin{figure}[!htbp]
\centering
\includegraphics{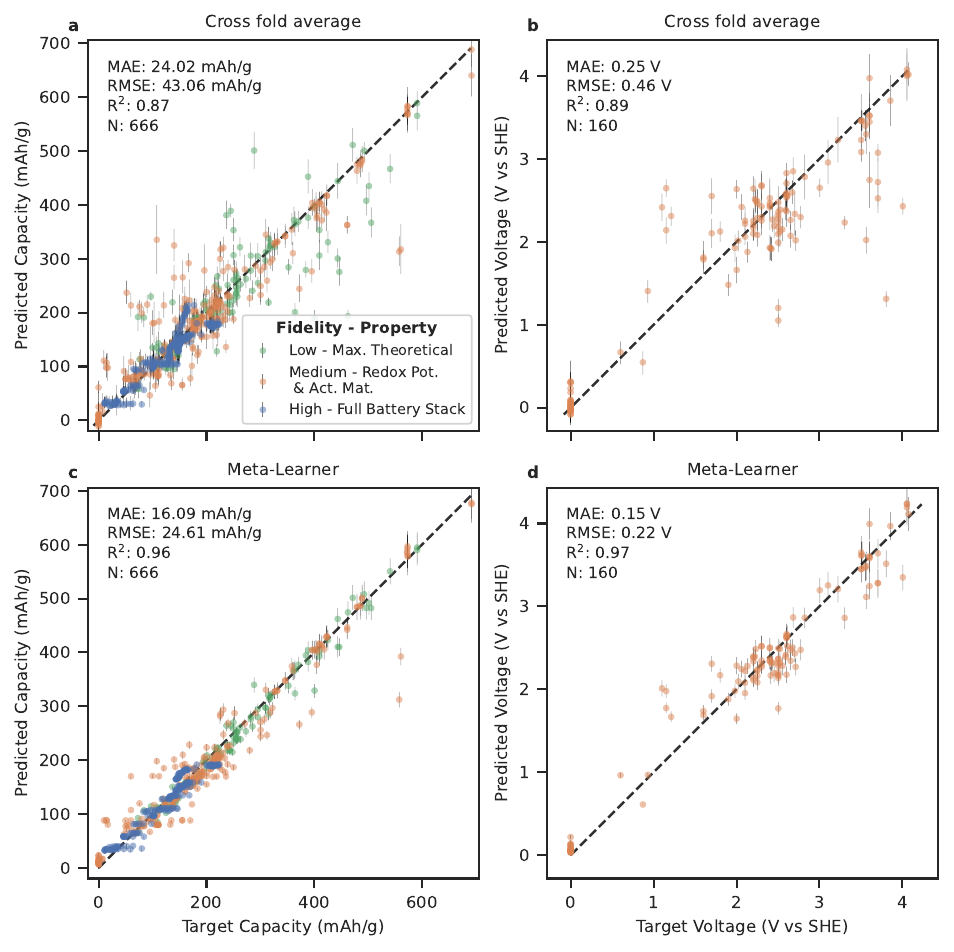}
\caption{Parity plots for HorizON and the meta-learner. Panels \textbf{a} and \textbf{b} show the five-fold averaged HorizON predictions for specific capacity and redox potential (V vs.\ SHE). Panels \textbf{c} and \textbf{d} show the corresponding meta-learner predictions. Points are colored by fidelity level. Low-fidelity points represent theoretical capacities. Medium-fidelity points represent DFT-computed redox potentials and experimental measurements without usage conditions or electrolyte information. High-fidelity points represent full battery-stack measurements. Each panel reports MAE, RMSE, $R^2$, and the number of test points $N$.}
\label{fig:capacity}
\end{figure}

\paragraph{Model performance}
In HorizON, we train five models using five-fold cross-validation subset, then ensemble their predictions using a meta-learner. Figure~\ref{fig:capacity}\textbf{a} and Figure~\ref{fig:capacity}\textbf{b} present parity plots of the fold-averaged predictions. The error bars represent two standard deviations ($2\sigma$) across the five models. HorizON achieves $R^2$ values of \num{0.87} for specific capacity and \num{0.89} for voltage.

The points in Figure~\ref{fig:capacity}\textbf{a} and Figure~\ref{fig:capacity}\textbf{b} are colored by fidelity level. The high-fidelity subset has a maximum specific capacity of \qty{223}{\milli\ampere\hour\per\gram}. These measurements describe complete battery stacks under defined usage conditions and therefore reflect practical electrode behavior. Practical behavior includes the influence of electrochemically inactive binders and conductive agents, together with capacity losses caused by dissolution and side reactions. In contrast, the medium-fidelity active-material capacities range from \qtyrange{8}{691}{\milli\ampere\hour\per\gram}, while the low-fidelity theoretical capacities range from \qtyrange{93}{570}{\milli\ampere\hour\per\gram}. These subsets omit the complete electrolyte and usage condition information or rely on theoretical capacity calculations. They therefore provide more idealized estimates of active-material performance. The richer battery-stack information produces tighter predictions for the high-fidelity targets. Joint training across all fidelity levels allows HorizON to retain this accuracy despite the limited number of high-fidelity measurements.

Reference values close to zero appear in all four panels of Figure~\ref{fig:capacity}. These points correspond to insulating materials included in the medium-fidelity capacity and experimental-voltage datasets. Their poor electronic conductivity produces negligible electrochemical capacity and voltage responses. This allows HorizON to distinguish between redox active and non-redox active materials. 

For the final predictions, we construct a meta-learner as an ensemble of the five fold HorizON models described in the Methods. For each battery stack, the five HorizON predictions are the inputs to the meta-learner that predicts the final property values. Figure~\ref{fig:capacity}\textbf{c} and Figure~\ref{fig:capacity}\textbf{d} show that the meta-learner has $R^2$ values of \num{0.96} for specific capacity and \num{0.97} for redox potential. The resulting predictions lie closer to the parity line and exhibit reduced scatter and smaller error bars. This improvement extends across all fidelity levels, including the high-fidelity battery-stack data. Performance metrics for each property and fidelity level are reported in Supplementary Table~S1. HorizON therefore learns cycle-dependent and scalar targets within a shared representation, while the meta-learner reduces variability among the fold-specific models.

\FloatBarrier

\paragraph{Inverse Design}
\begin{figure}[!htbp]
\centering
\hspace*{-1cm}%
\includegraphics{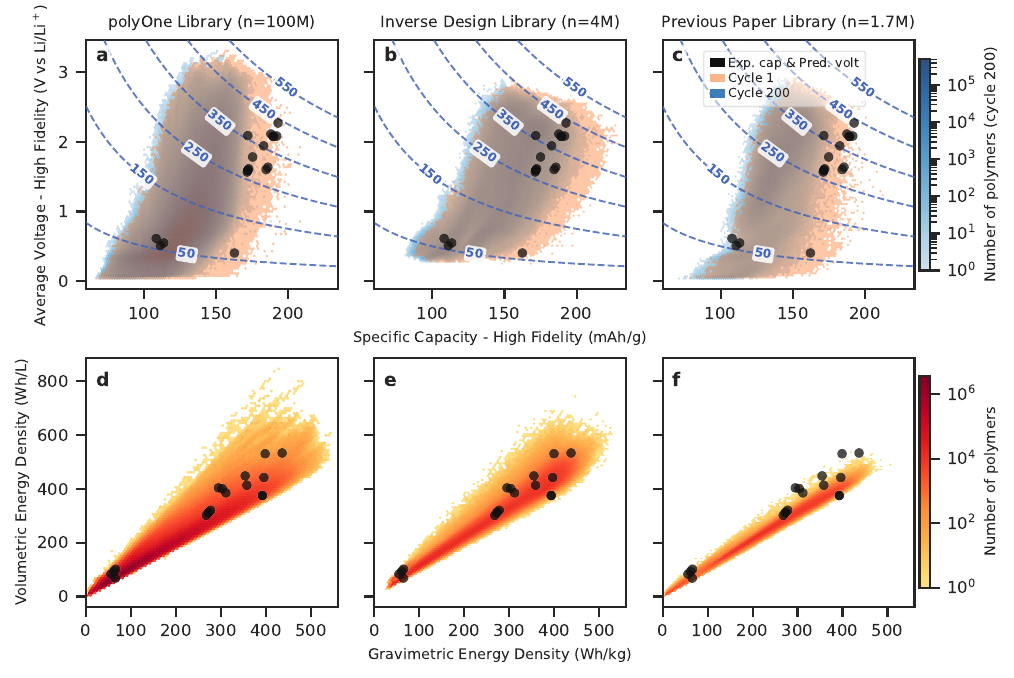}
\caption{Specific capacity versus average voltage (panels \textbf{a} to \textbf{c}) and
gravimetric versus volumetric energy density (panels \textbf{d} to \textbf{f}) for three
polymer libraries, predicted at high fidelity for the LiTFSI/TEGDME
electrolyte at a 1C rate for the \nth{1} and \nth{200} cycles. Voltage values are average working potentials versus lithium
(V vs.\ \ce{Li+}/Li). Panels \textbf{a} and \textbf{d}:~The public
polyOne library (\qty{100}{M}).\cite{christopher_kuenneth_2022_7766806}
Panels \textbf{b} and \textbf{e}:~The inverse-design library generated in this work
(\qty{4}{M}). Panels \textbf{c} and \textbf{f}:~A single-cycle library from our previous work
(\qty{1.7}{M}).\cite{Ganti2025} Gravimetric energy density is the product of
capacity and voltage; volumetric energy density additionally uses a
computationally estimated polymer density. The color scale
gives the number of polymers per bin and is shared across all six panels.
Dashed lines in panels \textbf{a} to \textbf{c} are energy-density contours
(\unit{\watt\hour\per\kilogram}). Blue points denote the experimental
cathodes used to train the property predictors: capacities are measured
values, voltages are model predictions, and volumetric energy
densities use the same estimated density as the library predictions.}

\label{fig:propspace}
\end{figure}

\begin{figure}[!htpb]
\centering
\includegraphics{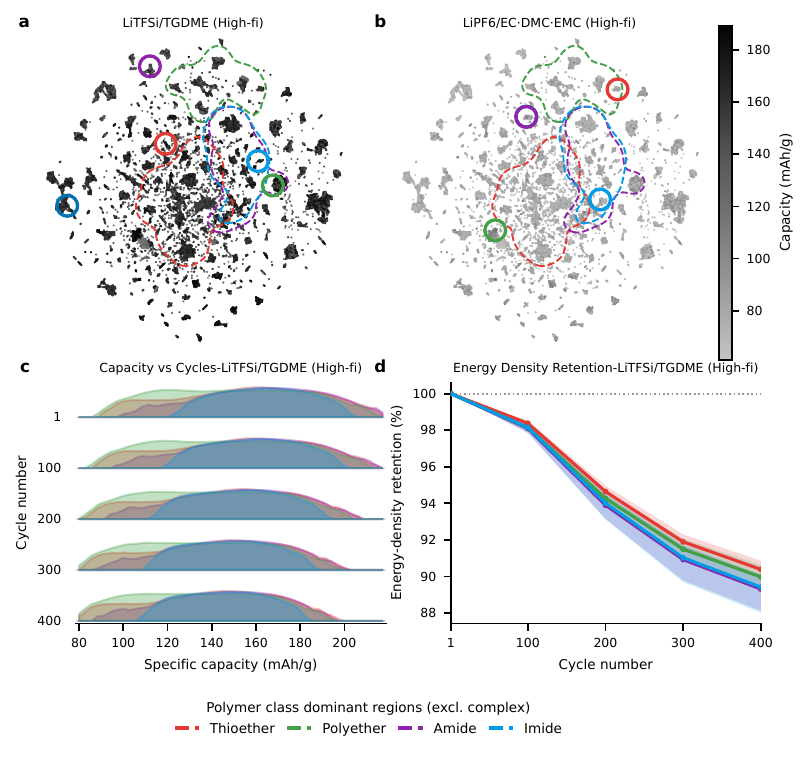}
\caption{UMAP projections and cycling predictions at high-fidelity for our inverse-design library of
\qty{4}{\million} polymers. Panels \textbf{a} and \textbf{b} show the predicted
specific capacity for the LiTFSI and LiPF$_6$ electrolytes. Dashed outlines mark the dominant region of
each polymer class in the UMAP plot, and solid circles mark the top candidate of each
class. Panel \textbf{c} is a ridgeline plot of the
library-wide specific-capacity distribution at cycles \numlist{1;100;200;300;400}
for LiTFSI/TEGDME; the distributions shift toward lower capacity as cycling proceeds.
Panel \textbf{d} shows the predicted energy-density retention (\unit{\percent}) of
each polymer class over \num{400} cycles. }
\label{fig:inverse}
\end{figure}

We utilize the trained meta-learner for inverse design. Starting from the polymer chemistries in our dataset, we iteratively substitute redox-active moieties and fragments along the main and side chains to generate a library of four million candidates. We evaluate every candidate at high fidelity and rank the candidates by predicted energy density within each polymer class. The complex polymer class denotes candidates with several redox-active moieties on one backbone.

Figure~\ref{fig:propspace} shows the high-fidelity property space of the three
libraries for the LiTFSI/TEGDME electrolyte at \num{1}C at the \nth{1} and
\nth{200} cycles. Across all three libraries, the \nth{200}-cycle predictions
shift toward lower specific capacities relative to the \nth{1}-cycle predictions,
while preserving the overall capacity--voltage trends. Figure~\ref{fig:propspace}a shows the
polyOne library, a general purpose dataset of \qty{100}{\million} polymers. Its property space spans a broad region of the capacity--voltage plane and includes a large low-voltage population. Figure~\ref{fig:propspace}\textbf{b} shows the inverse-design library generated from the existing polymer materials in the current work. Figure~\ref{fig:propspace}\textbf{c} shows the polymer library generated in our previous work and reused here\cite{Ganti2025}. Although their property spaces overlap, the current inverse-design library extends further toward high specific capacities at intermediate voltages. Many newly generated polymers in this library have predicted specific capacities above the measured capacities of the literature cathodes shown as blue points in Figure~\ref{fig:propspace}\textbf{b}. The library from our previous work occupies a narrower capacity range across much of the voltage axis and shows a more prominent band near zero voltage. Both libraries are evaluated here at high fidelity under the same battery-stack and usage conditions, including the conductive agent, binder, and electrolyte. The differences between Figure~\ref{fig:propspace}\textbf{b} and Figure~\ref{fig:propspace}\textbf{c} therefore compare the two polymer libraries rather than different prediction fidelities. In every panel the densest
region follows the intermediate energy-density contours. The upper-right corner
stays sparse exhibiting the voltage-capacity tradeoff because only a few polymers reach
high capacity and high voltage together. The predicted energy
densities extend to roughly \qty{550}{\watt\hour\per\kilo\gram}. Figure~\ref{fig:propspace}d--f show the same three libraries as
gravimetric versus volumetric energy density. Volumetric energy density
is calculated using polymer densities estimated from Wildman--Crippen molar
refractivity and the Lorentz--Lorenz relation (see Methods). Each library
occupies a relatively narrow region in this energy-density space. The
experimental cathodes lie within these regions. Their positions
in Figure~\ref{fig:propspace}d--f agree with Figure~\ref{fig:propspace}a--c: candidates with favorable capacity
and voltage also show favorable estimated volumetric energy density.
The highest-ranked candidates contain carbonyl or quinone redox cores on extended
conjugated backbones. They also carry thioether sulfur, ring nitrogen, or
unsaturated linker end groups, as supported by previous
studies~\cite{Raj2021,Xu2023,Hatakeyama-Sato2023}.

Figure~\ref{fig:inverse} shows the high-fidelity landscape of the inverse-design
library produced in this work. Panels \textbf{a} and \textbf{b} show the UMAP
plot of predicted specific capacity. Our model predicts higher capacities
with LiTFSI/TEGDME than with LiPF$_6$ across the entire embedding which agrees with reports that ether-based electrolytes suppress dissolution of organic active materials~\cite{shi_challenges_2022,lee_recent_2018}. The top candidates reach
\qty{227}{\mAh\per\gram} with LiTFSI/TEGDME and \qty{143}{\mAh\per\gram} with
LiPF$_6$ (see Supplementary Figures S1--S2). The top candidates occupy different UMAP regions rather than regions dominated by a single polymer class. This arises because their multiple redox-active moieties produce fingerprints that lie between the single-class regions. The top candidates consists of a carbonyl or quinone redox core with a fused aromatic conjugated framework. Incorporating thioether sulfur or ring nitrogen provides an additional advantage. The highest predicted oxidation and reduction potentials occur for thianthrene- and phenothiazine-type sulfur aromatics and hindered pyridine-ether esters, consistent with the literature (Supplementary Figures S3-S4)~\cite{Hatakeyama-Sato2023}.

Figure~\ref{fig:inverse}c and Figure~\ref{fig:inverse}d summarize the predicted
cycling response of the LiTFSI/TEGDME candidates. Figure~\ref{fig:inverse}e reports
the library-wide specific-capacity distribution at successive cycle numbers. The
class distributions overlap over a common capacity range and shift toward lower
capacity as cycling proceeds, with a predicted average loss of
\qtyrange{30}{40}{\mAh\per\gram} over \num{400} cycles. Figure~\ref{fig:inverse}f
reports the predicted energy-density retention of each polymer class over the same
interval. All classes retain a comparable fraction of their initial energy density,
between about \qtyrange{89}{91}{\percent} after \num{400} cycles, and the retention
curves overlap within their confidence bands.

\begin{figure}[!htbp]
\centering
\includegraphics{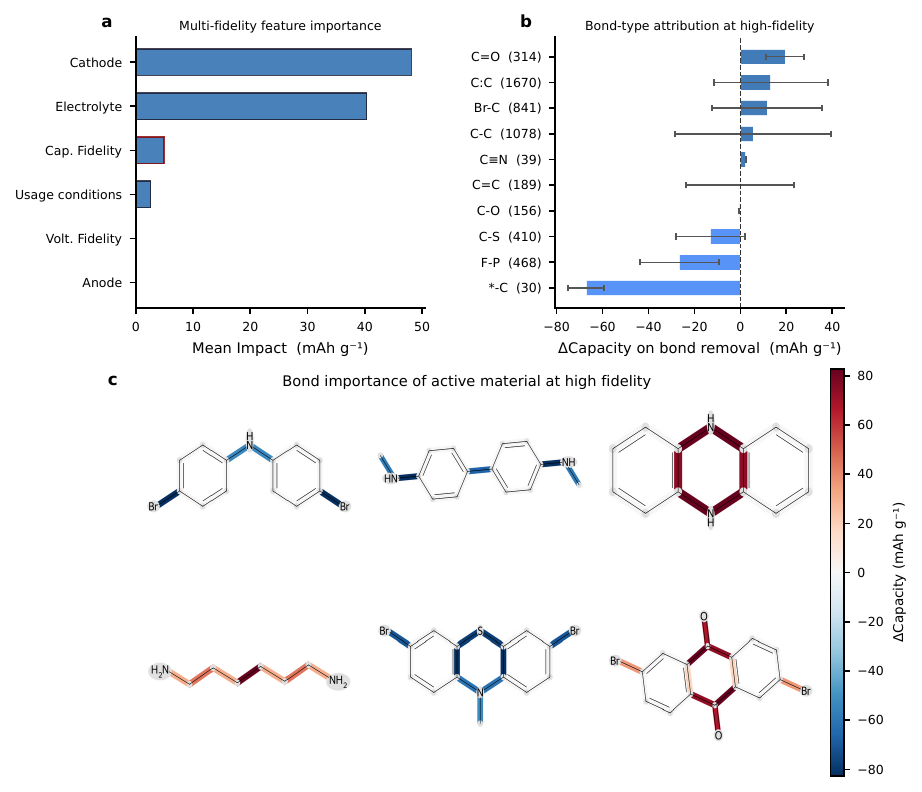}
\caption{SHapley Additive exPlanations (SHAP) analysis of the specific-capacity predictions. Panel \textbf{a} shows the mean impact of each input component group across all fidelity levels. Panel \textbf{b} shows the mean change in high-fidelity predicted capacity when a given bond type is removed from the cathode active material, with the bond counts given in parentheses. The error bars denote the spread across the sampled polymers. Panel \textbf{c} maps the high-fidelity per-bond capacity attribution to six monomers, which are used as active materials of cathode copolymers. The colors range from capacity-reducing (blue) to capacity-enhancing (red).}
\label{fig:inverse-results}
\end{figure}

\paragraph{SHAP Analysis}
We use SHapley Additive exPlanations (SHAP)\cite{Lundberg2017} to quantify contributions to predicted specific capacity. We perform the analysis on the five cross-validated models at two levels: battery-component groups and chemical bonds within the cathode active material (Figure~\ref{fig:inverse-results}). The component-level analysis in Figure~\ref{fig:inverse-results}a includes capacity predictions from all fidelity levels. The cathode and electrolyte dominate the predicted capacity, whereas the anode contributes minimally because all stacks use the same lithium-metal anode. The capacity-fidelity inputs encode the fidelity of the capacity labels and contribute slightly, whereas the voltage-fidelity inputs contribute negligibly. The large electrolyte contribution confirms that capacity is a battery-stack property governed by dissolution and interfacial chemistry \cite{Zhang2025,Yang2025}. This result supports training on the full battery stack rather than the cathode alone. Moreover, the SHAP analysis strategy evaluates the model and adds to its credibility

For the chemical fragment analysis of the cathode active material (Figure~\ref{fig:inverse-results}b and Figure~\ref{fig:inverse-results}c), we focus on high-fidelity predictions that include all battery components. Chemical fragments are ranked by the mean decrease in predicted capacity when the corresponding bond is removed from the cathode active material (Figure~\ref{fig:inverse-results}b). C=O, C:C, and Br--C bonds show the largest positive contributions to specific capacity. The strong attribution to carbonyl (C=O) bonds is consistent with their established role as redox centers in organic cathodes. On the other hand, C--C and imine (C$=N$) bonds show weaker impact on capacity and are associated with larger uncertainty, reflecting the sensitivity of these fragments to the local chemical environment\cite{Guo2025}. Moreover, F--P bonds are associated with reduced specific capacity, which is due to a higher thermodynamic barrier for electroactivity\cite{Hautier2011}. These trends are consistent with prior literature \cite{Peng2019,Wu2015,Shu2024,Liu2024,Zhu2019}. Figure~\ref{fig:inverse-results}c maps the chemical fragment attribution on the six monomers that are utilized as the cathode active materials. Carbonyl groups and nitrogen-containing conjugated heterocyclic backbones are generally capacity-enhancing (red), whereas brominated carbon and nitrogen in non-conjugated structures are neutral or capacity-reducing (blue). Collectively, the results suggest design guidelines in which redox-active carbonyl groups and conjugated backbones support higher capacity, whereas non-conjugated structures with multiple heteroatoms tend to exhibit lower capacity.

\section{Discussion}
Here, we introduce HorizON, a general predictor of organic-battery stack properties trained across multiple properties, fidelity levels, and time horizons. HorizON can predict cycle-resolved specific-capacity trajectories at high fidelity for full stacks, using fingerprints of the cathode, electrolyte, and lithium-metal anode. HorizON is trained in a multi-task, multi-fidelity setup that connects these high-fidelity trajectories with medium-fidelity DFT redox potentials and lower-fidelity proxy targets. This joint training helps the model capture correlations across properties and fidelities, allowing information learned from one task to improve another. It also mitigates data scarcity by combining datasets even when some datapoints lack certain property and fidelity labels. After training, HorizON makes predictions in under \qty{1}{\milli\second} per stack.

The space of possible organic battery stacks is extremely large because performance depends on the choice of stack components, their compositions, and usage conditions. By providing rapid, stack-level predictions of high-fidelity capacity trajectories, HorizON can support larger-scale exploration of this space and help prioritize cycling experiments. As with any data-driven model, its accuracy is bounded by the chemistries and cell designs represented in the training data. In particular, our high-fidelity stacks use lithium metal as the anode, so extending to other anodes, full-cell configurations, ion systems, or broader electrolyte families will require additional data and experimental validation. In addition, the inverse-design workflow does not evaluate synthetic accessibility, polymerizability, repeat-unit stability, or electrode-fabrication compatibility, so proposed candidates should be confirmed by synthesis and electrochemical testing. Looking forward, multi-task, multi-fidelity models like HorizON can also be used for knowledge transfer across related battery domains and to guide targeted data collection for rapid model adaptation.

\section{Methods}

\paragraph{Data collection and curation}
We collect and curate a multi-fidelity dataset with \num{1068} data points spanning experimental cycling measurements, first-cycle experimental properties, theoretical capacities, and DFT redox potentials. We extract the high-fidelity cycling subset from peer-reviewed publications and review articles on organic cathode half-cells with lithium-metal anodes. For each experimental system, we collect specific capacity as a function of cycle number and the corresponding usage conditions with some trajectories extending to approximately \num{400} cycles. For each cathode formulation, we collect PSMILES strings of the active organic material and SMILES strings for conductive agent and binder. For each electrolyte, we collect the SMILES strings of the salt and solvent. For the DFT subset, we calculate redox potentials of the redox-active component in different oxidation and reduction states. Before splitting the data into training and validation datasets, we apply min--max scaling independently to each target property.

\begin{figure}[H]
\centering
\includegraphics[width=0.5\columnwidth]{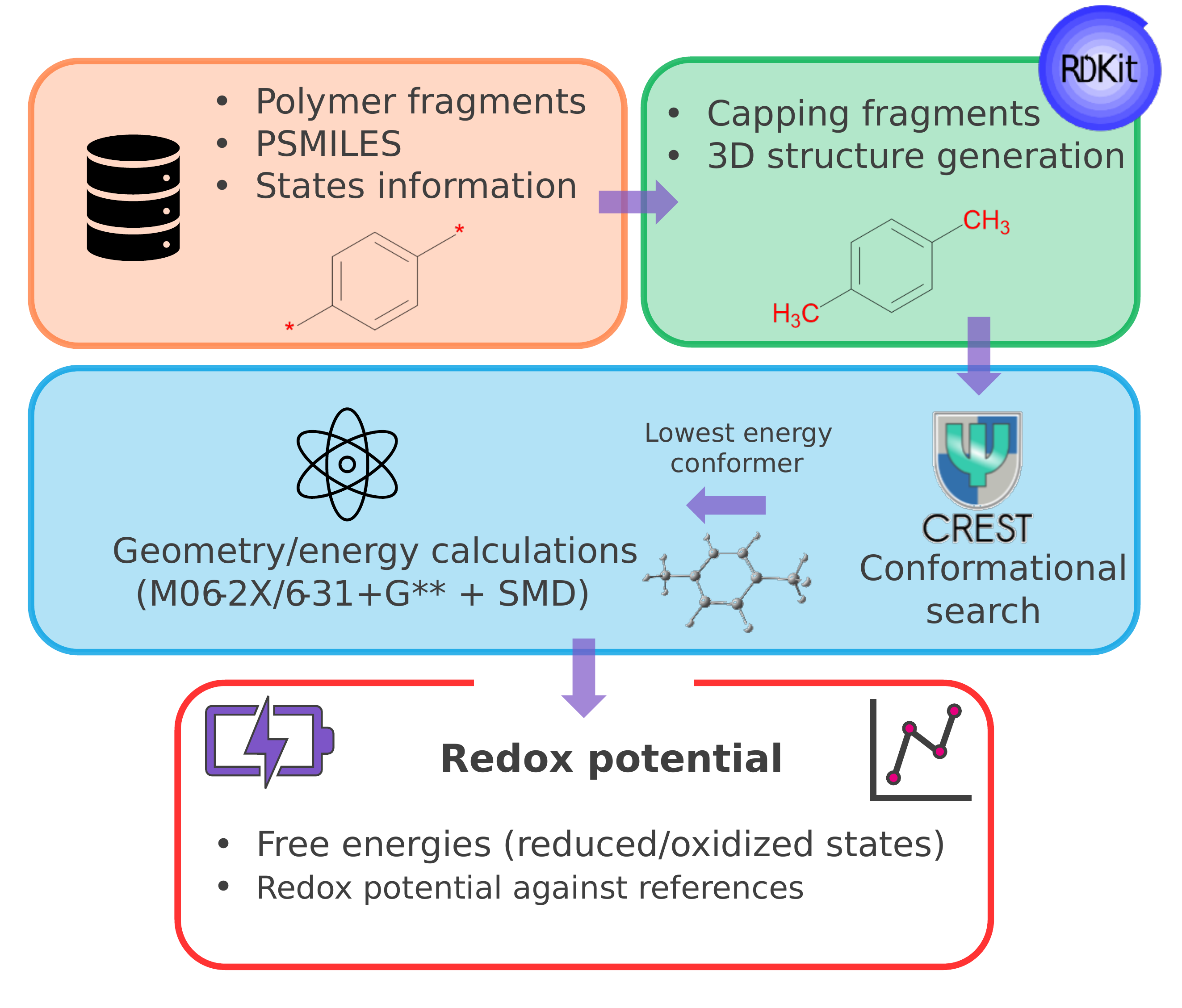}
\caption{Schematic of the automated DFT redox potential workflow. Starting from PSMILES strings, the pipeline proceeds through methyl capping and 3D structure generation (RDKit), conformer search (CREST/GFN2-xTB), geometry optimization (PCM-M06-2X/6-31+G**), and final single-point energy evaluation (SMD-M06-2X/6-31+G**) to yield redox potentials referenced to SHE.}
\label{fig:dft_workflow}
\end{figure}

\paragraph{DFT Redox Potential Calculations}
An automated computational workflow was developed that takes PSMILES strings and predicts redox potentials (see Figure~\ref{fig:dft_workflow}). The workflow starts by capping redox-active polymer fragments with methyl groups and generating 3D structures using RDKit\cite{greg_landrum_2025_15286010}. A conformer search is then performed using the CREST program~\cite{Pracht2024} and GFN2-xTB method~\cite{Bannwarth2019}, automatically selecting the lowest-energy conformer for further optimization. This structure is re-optimized using the M06-2X functional~\cite{Zhao2007} with 6-31+G** basis sets~\cite{Nobes1982} in a PCM solvation model~\cite{TOMASI1999211} with acetone as the solvent. The dielectric constant is \num{20.7},
which represents the average of the experimental solvents~\cite{Hall2015}. Previous studies~\cite{PrataDuque2022} have shown that DFT method choice does not significantly impact these results. For the final free energies, the calculations switch to the SMD solvation model~\cite{Marenich2009} (still with acetone) to obtain more accurate solvation energies, yielding liquid-phase free energies for both reduced and oxidized states at the SMD-M06-2X/6-31+G**//PCM-M06-2X/6-31+G** level. Redox potentials are derived from the free-energy difference between oxidized and reduced states via a standard thermodynamic cycle and referenced to SHE. All stationary points are confirmed as true minima by frequency analysis. The workflow interfaces RDKit, CREST, and Gaussian~16\cite{g16} to enable systematic processing of large polymer candidate sets.

\paragraph{Fingerprinting}
We encode each SMILES or PSMILES representation as a \num{600}-dimensional polyBERT fingerprint \cite{kuenneth_polybert_2023}. For multicomponent materials, the composition-weighted fingerprint is computed as
\begin{equation}
\mathbf{f}_{\mathrm{mix}} = \sum_{i=1}^{N} w_i \, \mathbf{f}_i,
\end{equation}
where $w_i$ and $\mathbf{f}_i$ denote the fraction and fingerprint of constituent $i$, and $N$ is the number of constituents in a multicomponent material\cite{kuenneth_copolymer_2021}. The cathode, anode, and electrolyte vectors are concatenated with seven selector variables: C-rate, active-material loading, electrolyte molarity, lower and upper voltage limits, oxidation state, and reduction state. This procedure produces the \num{1807}-dimensional model input.

\paragraph{Multi-Horizon Deep Operator Network (HorizON) Architecture}

We split the dataset into a cross-validation (\qty{80}{\percent}) and a meta-learner subsets (\qty{20}{\percent}). We divide the cross-validation subset into five folds. For each fold, Optuna optimizes the number of layers, neurons per layer, activation functions, and learning rate\cite{akiba_optuna_2019}. We select the best model for each fold based on the validation performance.

Each of these cross-validated models predicts the property values for every sample in the meta-learner subset. The five predictions for each sample form a \num{5}-dimensional input vector. We train the meta-learner to predict the target property from this \num{5}-dimensional vector. The complete prediction framework, including the five cross-validated models and the meta-learner, constitutes HorizON. During inference, a new sample is passed through the five cross-validated models, and their predictions are supplied to the meta-learner to obtain the final property prediction.

\paragraph{Inverse Design}
Our inverse-design workflow starts from a set of \num{65} homopolymers drawn from the training dataset. We use these polymers as the starting polymer chemistries to generate a large virtual library of material candidates. We generate the library iteratively by substituting selected positions in the PSMILES strings with heteroatoms and related chemical modifications. After each substitution step, we evaluate chemical validity using RDKit and PSMILES-based checks \cite{greg_landrum_2025_15286010,kuenneth_polybert_2023}. This procedure yields approximately \num{4000000} chemically valid polymer candidates. We compute their polyBERT fingerprints and predict their properties using the trained meta-learner. Candidates with the highest predicted redox potentials and specific capacities are reported in the Supporting Information.

\paragraph{Volumetric \& Gravimetric Energy Density Calculations}

The computation of the volumetric energy density requires the mass density of each polymer. We estimated the density for each polymer's repeat-unit SMILES string using the
RDKit's implementation of the Wildman--Crippen atomic-contribution model\cite{Wildman1999}, which sums per-atom contributions to molar
refractivity ($MR$) from a fitted set of atom types. We converted $MR$ to
molar volume ($V_m$) using the classical Lorentz--Lorenz relation,
\begin{equation}
V_m = MR\frac{n^2+2}{n^2-1}, \qquad n \approx 1.55,
\end{equation}
where $n$ is the refractive index and \num{1.55} is typical of organic
polymers~\cite{VanKrevelen2009}. The density is
\begin{equation}
\rho = \frac{M}{V_m},
\end{equation}
where $M$ is the molecular weight of the repeat unit. To obtain the repeat-unit SMILES, the end-group markers (represented by $*$) in PSMILES are replaced by hydrogen atoms. Gravimetric energy density followed as
\begin{equation}
E_\mathrm{grav} = 3.6\times10^{-3}\,Q\,V,
\end{equation}
where $Q$ is capacity in
\unit{\milli\ampere\hour\per\gram} and $V$ is voltage in volts, giving
$E_\mathrm{grav}$ in \unit{\mega\joule\per\kilogram}. Volumetric energy
density followed as
\begin{equation}
E_\mathrm{vol} = E_\mathrm{grav}\rho.
\end{equation}
Because $\rho$ is a computational estimate rather than a measured or
literature value, we report volumetric energy density as a model
estimate throughout.

\paragraph{SHAP Analysis}
In the current work, we utilize a SHAP analysis strategy to quantify the importance of local chemical motifs. We estimate how individual bonds in the cathode active material contribute to the predicted capacity of the battery stack. We systematically remove individual bonds from the PSMILES strings of the cathode active material and generate polyBERT embeddings for the modified structures. We use these embeddings as inputs to the five cross-validated models and calculate the change in predicted capacity relative to the original prediction. The resulting attribution score measures the contribution of the removed bond to the predicted property. Larger positive values indicate a stronger contribution to the predicted capacity.

\section{CO$_2$ Emission and Timing}
We estimate the training emissions from the Bayreuth cluster carbon intensity of \qty{0.344}{\kilogram\COtwoeq\per\kWh}. A \qty{40}{\hour} run on \num{4} A100-\qty{40}{\giga\byte} GPUs with a \qty{250}{\watt} TDP corresponds to approximately \qty{3.44}{\kilogram\COtwoeq}. We estimate inference emissions as \qty{6.5}{\gram\COtwoeq} per candidate for a \qty{2}{\second} prediction.

\section{Data Availability}
Data are available upon request.
\section{Code Availability}
The code used for training the HorizON models and meta learner models are available for academic use at \url{https://github.com/kuennethgroup/HorizON.git}.
\bibliography{references}
\section{Acknowledgements}
S.V.S.G., L.W, A.D, and C.K acknowledge funding from the BayBatt graduate school. Additionally, T.M, T.N, and J.M would like to express their gratitude to Jens Langanke, Head of R\&D Simulations \& Data Science, Covestro for valuable insights and for backing initiatives associated with this project.

\section{Author Contributions}
S.V.S.G.: conceptualization, data, models, analysis. L.W and A.D.: data curation. T.M, T.N and J.M.: computational methodology, automated workflow development, DFT redox potential calculations, and computational data curation. C.K.: conceived, guided and funding acquisition. All the authors discussed results and commented on the manuscript. 

\section{Competing Interests}
None declared.

\end{document}